\documentclass[manuscript]{aastex}

\shorttitle{}
\shortauthors{}

\begin{document}

\title{Formation of sharp edges and planar areas of asteroids by polyhedral
abrasion}

\author{G. Domokos\altaffilmark{1,2}, A. \'A. Sipos\altaffilmark{1}, Gy. M.
Szab\'o\altaffilmark{4}, and P. L. V\'arkonyi\altaffilmark{1}}
\altaffiltext{1}{Department of Mechanics, Materials, and Structures,
Budapest University of Technology and Economics, M\"uegyetem rkp. 3,
Budapest 1111, Hungary} \altaffiltext{2}{Trinity College, Cambridge, CB2
1TQ} \altaffiltext{3}{Department of Experimental Physics, University of
Szeged, D\'om t\'er 9. Szeged 6720, Hungary} \altaffiltext{6}{E-mail:
gd321@cam.ac.uk, szgy@titan.physx.u-szeged.hu}

         \begin{abstract}
While the number of asteroids with known shapes has drastically
increased over the past few years, little is known on the
 the time-evolution of shapes and the underlying physical processes.
Here we propose an averaged abrasion model based on micro-collisons,
accounting for asteroids
not necessarily evolving  toward regular spheroids, 
rather (depending on the fall-back rate of ejecta)
following an alternative path, thus confirming 
photometry-derived features, e.g. existence of large,
relatively flat areas separated by edges.  We show
that our model is realistic, since the bulk of the collisions
falls into this category.
         \end{abstract}


                 \section{Introduction}

The number of asteroids with known shapes has increased significantly in
the past years. 8 asteroids were visited by fly-by missions, providing
high-resolution images (e.g. Veverka et al. 1994, Chapman et al. 1995,
Robinson et al. 1995, Veverka et al. 1999, Oberst et al. 2001, Duxbury et
al. 2004, Saito et al. 2006, Schulz et al. 2008). Radar observations can
also reveal the envelope of the shape (e.g. Ostro et al. 1988). A very
promising tool of shape identification is the inversion of photometric
light curves (Kaasalainen and Troppa, 2001, Kaasalainen et al 2002,
KTP2002 hereafter), which resulted in photometric models for $\approx$100 asteroids, so far.
These new efforts and techniques lead
to the rapid increase of the number of known asteroid shapes and we can
expect this number to keep growing in the future.

These shapes have not been globally classified in the literature, but one
can easily recognize some typical shape properties. On one hand, there are
some asteroids with highly nonconvex shape (e.g. 3908 Nyx), there are
bifurcated objects (Kleopatra), earthnut-like forms (Eros), all of which
may be results of typical formation processes. On the other hand, several
asteroids have a strikingly simple shape envelope with large flat surfaces
bordered by relatively sharp edges. A 
significant portion of photometric shapes
apparently share these features (e.g. 6053
{1993 BW}, \v{D}urech et al. 2002, 10115
1992 KS, Busch et al. 2005, 1580 Betulia, 1980 Tezcatlipoca, 2100
Rha-Shalom, Kaasalainen et al. 2004). These observations are often attributed to
local concavities (e.g. \v{D}urech 2002) or albedo effects (KTP2002), i.e. they
have been claimed to be artefacts of the
photometric method. However, close-up images from spacecrafts also reveal
that the presence of prominent flat areas and sharp edges is typical for
at least half of the small asteroids (e.g. Ida, Gaspra, Annefrank, Steins).
Recently, Troppa et al (2008) confirmed
the presence of flat areas on photometric {shapes}
and found that the number of flat areas/facets decreases
with the elongation of the shape (or its convex hull).
While photometric imaging certainly does not provide sufficient
information to quantify the uncertainities associated with the
mentioned qualitative geometric features, these images suggest that
at least some asteroids may not evolve towards spheroids. Our goal
is to explore this avenue by applying a simple averaged abrasion model.

Recently, several studies have been dedicated to explain the shapes and
surface properties. The conclusions point to different evolution scenarios; 
they also help 
to better understand the relation between collisions
and the global shape. The following basic scenarios were proposed:

         \begin{itemize}
         \item{}Formation of primordial shapes in {disrupting,} energetic collisions
(Michel et al 2003), followed by continuous formation and erosion of craters
(Housen and Holsapple, 2003, O'Brien et al. 2006).
         \item{}Rubble pile structure asteroids can get very elongated
forms as a result of tidal effects (Bottke et al. 1999).
         \item{}Evolution of rubble piles with elongated shapes toward a
spherical form by seismic shaking of
         collisions with moderate energy ( Korycansky and Asphaug, 2003, KA2003
	 hereafter, Szab\'o and Kiss, 2008)
         \item{}Continuous alteration of the surface by low-energy
collisions and space weathering
that form small craters (Lazzarin et al. 2006).
         \end{itemize}

The widespread concept is that small impactors 
always smoothen the global shape by eroding sharp edges (A. Conrad, personal
communication). The root of this concept is the obvious $O(3)$-symmetry of
the problem: the impacts have uniform radial intensity and one would expect
that abrasion in this environment converges to an $O(3)$-symmetric object
(sphere).

Low-energy impactors remove small pieces from the
surface of the asteroid, which either fall back to the surface 
(captured by the gravitational field), or they escape.
A model based exclusively on the first scenario (KA2003) predicted convergence 
to spheres
 (or oblate ellipsoids for rotating asteroids). In this paper, we adress the second one. 
In contrast to natural expectations, abrasion by small, escaping 
fragments breaks the $O(3)$-symmetry, 
and leads to the formation of large planar areas and sharp edges. 
The global outcome for most initial shapes are low-order convex polyhedra 
(i.e. convex shapes bordered by a few planes), similar to at least a 
considerable fraction of asteroids.
 We also present preliminary results for a more realistic model, 
combining the effects of escaping
 material and the redistribution of ejecta on the asteroid's surface.
 The reported features appear to preserved in a certain parameter-range.

                         \section{Description of the model}

Abrasion due to micro-impacts is a complex,
stochastic process, in which minor, discrete pieces are removed from the
asteroid's surface in small collisions. Our model is an averaged continuum
approach where the effect of large  numbers of small, discrete impactors is
modeled by continuous abrasion of the surface. The first step towards the construction
of the partial diffeential equation (PDE) is determining the
average rates of abrasion on an asteroid's surface under the simplest 
assumptions for the discrete impactors (e.g. uniform
distribution for the direction).

Let the asteroid surface at time $t$ be given by the endpoints of a
two-parameter family of vectors $\textbf{x}(p,q,t)$. We assume that a large
number of small particles (relative to the asteroid) move along random
straight lines (in coordinates fixed to the asteroid) and collide with the
asteroid; subsequently they escape together with small, broken portions
of the asteroid (Ronca and Furlong 1979). 
In case of impact trajectories we ignore gravitational effects
as well as spinning; both are negligible at high-speed impacts. 
(The role of gravity, however, can not be neglected when
considering material falling back to the surface; 
the model we first discuss assumes that all material escapes, later in the paper we develop
a general model including non complete ejecta escape.)

In particular, we assume
constant and uniform (i.e. time- space- and direction-independent) radial
intensity $\Phi$ of particle collisions (average number of collisions per
unit time, unit area and unit solid angle), resulting, in case of convex
surfaces, in constant and uniform surface intensity $m$ of particle
collisions (average number of collisions per unit time, and unit surface
area). We assume that the volume of material abraded from unit surface
per unit time, or, equivalently, the abrasion speed $\partial
\textbf{x}(p,q,t)/ \partial t$ in the direction of the inward directed
surface normal $\textbf{n}(p,q,t)$ is determined by the surface intensity
$m$ and the abrasion "efficiency" of the impactors. The latter could be
expressed as a function $f(\beta)$ of the impact angle $\beta$ ($0<\beta <
\pi/2$) between $\textbf{n}$ and the direction of the incoming particle.
Abrasion speed can be obtained by integrating $\Phi f(\beta)$ over the $HS$
half-sphere and considering that the number of the incoming particles
depend on $\beta$, %
         \begin{eqnarray}
         \frac{\partial \textbf{x}(p,q,t)}{\partial t} = \int_{HS}\Phi
\cos\beta
         f(\beta)dS\cdot \textbf{n}(p,q,t)\nonumber\\
         =\Phi
         \int_0^{\pi/2}f(\beta)\cos\beta\sin\beta d\beta \cdot\textbf{n}(p,q,t)
         \end{eqnarray}

In case of $f(\beta)=1$, Eq. 1 yields $|\partial \textbf{x}/ \partial t| =
\Phi = m$, however, for any other $f(\beta)$ we still obtain the simple
PDE called Eikonal equation for the abrasion of
convex surfaces due to uniform micro-impacts: %
         \begin{equation}
         \frac{\partial \textbf{x}(p,q,t)}{\partial t} = constant\cdot
\textbf{n}(p,q,t),
   \end{equation}

Of course, primordial asteroid shapes are often locally concave.
 In these regions equation (1) should be integrated
over its zenith angle, which is only a part of the half-sphere. Thus, the
surface intensity $m$ (and the abrasion speed) is  lower for concave surface
points than for typical (generic) points on the convex hull, and
therefore concave parts tend to disappear. In the rest of the paper, we
focus on the convex case.

After establishing the appropriate averaged continuum model, our next goal
is to show the resulting stable polyhedral geometric patterns.

Solutions of (2) are known as wave-fronts (Arnold, 1986), and
singularity theory describes their evolution,
Figure 1 illustrates the propagation of a planar
wave-front starting from a convex, smooth curve. 
One of the most relevant features is that
initially smooth surfaces evolve structurally stable singularities (cf.
Figure 1b) and self-intersections along edges(Figure 1c).
In a physical abrasion process, the solution is limited to the
domain free of self-intersections (which is a weak solution of the PDE,
(Sethian, 1985)): the aforementioned self-intersecting edges and vertices
appear on the surface of the physically relevant domain (Figure 1d). In
three dimensions, the situation is analogous: edges and vertices appear in
the solution for all generic (typical) smooth initial shapes (Figure 2a,b).
Structural stability means in this context that the appearance of these
geometric features can be expected for
almost all initial surfaces. Another feature of this process 
is that polyhedra evolve to polyhedra, with 
the number of faces monotonically decreasing in time (Figure 2c). These 
features imply that this process produces shapes with flat areas
separated by marked edges and vertices, almost regardless of the starting
configuration. One can also provide a simple and complete list
of final limit shapes (depending on the
initial shape): tetrahedron, elongated cigar-shape
or a flat disc (V\'arkonyi et al., 2008, Domokos et al, 2009.). We demonstrate two of these
limits numerically (Figure 2). Needless to say, more sophisticated discrete
 models would predict the process more accurately in the
quantitative sense, however, our
compact PDE model captures the most essential qualitative features.

\begin{figure}
\includegraphics[width=\columnwidth]{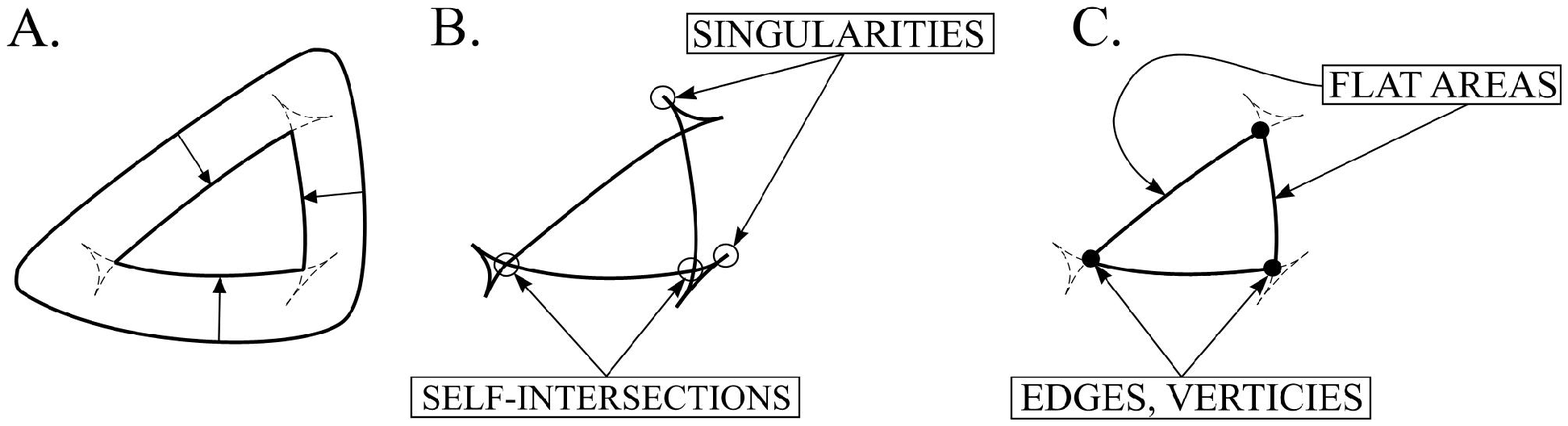}
\caption{Propagation of wave front inside planar curve. 1a:
Wave front before and after self intersection
1b: Generic (typical) singularities and self-intersections, occurring for
almost all initial curves 1c :Physically relevant part of wave front, with
edges resulting from self-intersections.}
\end{figure}

Previously, the evolution of asteroid shapes was modelled in two dimensions
by Ronca and Furlong (1979) under similar assumptions, however,
involving two misleading concepts: (i) rate of volume loss was associated with
\textit{radial} abrasion speed
(toward the barycenter) rather than speed normal to the surface, and (ii)
a false interpretation of zenith angle lead to nonconstant
abrasion speeds on convex surfaces and also to inconsistent results. Their
model also predicts surface singulaities,
however, otherwise radically different morphological characters then ours.

                         \section{Simulations}

Eq. (2) is a Hamilton-Jacobi type PDE, which can
be efficiently simulated by level-set methods (Osher and Fedkiw, 2002) and
fast-marching methods (Sethian, 1999). Both methods handle well singularities
(edges and vertices). In this paper, a MATLAB-based implementation
of the level-set method (Mitchell, 2008) was used to compute the evolution
of various initial shapes.

As reported from experiments
e.g. by Cappacioni et al. (1984), Catullo et al (1984) and
Ryan et al (2000), the average axis ratio of
asteroids resulting from fragmentation of larger objects is
1/1.4/2 with large scatter in both ratios. 
We used initial axis ratios even closer to the sphere to demonstrate
that the emergence of elongated shapes and edges is independent
of the initial shapes.

Figure 2 shows the evolution of three different random
initial shapes of which the first two have been
picked from relatively smooth surfaces. Observe emerging sharp edges and
plane areas, analogous to the planar wave front in Figure 1. 
Also, observe the tetrahedral limit in the first case and
the increasing prolateness leading to a 'cigar-shape' in the second case.
The third series starts
with a multi-faceted polyhedral shape (which could be interpreted as the
convex hull of a more general shape). Here
the number of faces is reduced in abrasion,  implying the emergence of low-order polyhedra
(and finally, a flat tetrahedron).
The emerging prolateness for many initial shapes 
confirms the observation of Troppa et al.
(2008). The figure also contains photometric shapes (Troppa et al 2002,
KTP2002, Kaasalainen et al. 2004) showing visual resemblance to the
intermediate stage of the simulated
examples. We emphasize, however, that it was not our goal
 to reconstruct the exact history of any
particular asteroid shape, we merely intended to illustrate the strong
qualitative agreement between 
photometric shapes and the simulated geometries.

All asteroid images shown in Figure 2 are the result of photometric
modeling providing only approximate information. 
We also tried  to reconstruct 
the evolution of one asteroid for which the shape
is known in detail. Asteroid 5535 Annefrank, visited by Stardust (Duxbury et al.
2004) apparently has two components, the major part is
similar to a tetrahedron. Figure 3a shows a hypothetical time-series
chosen by trial-and error from simulations of randomly perturbed spheres.
One stage of the evolution is a rounded polyhedron,
strongly resembling 5535 Annefrank (Fig 3c).

\begin{figure}
\begin{center}
\includegraphics[width=8cm]{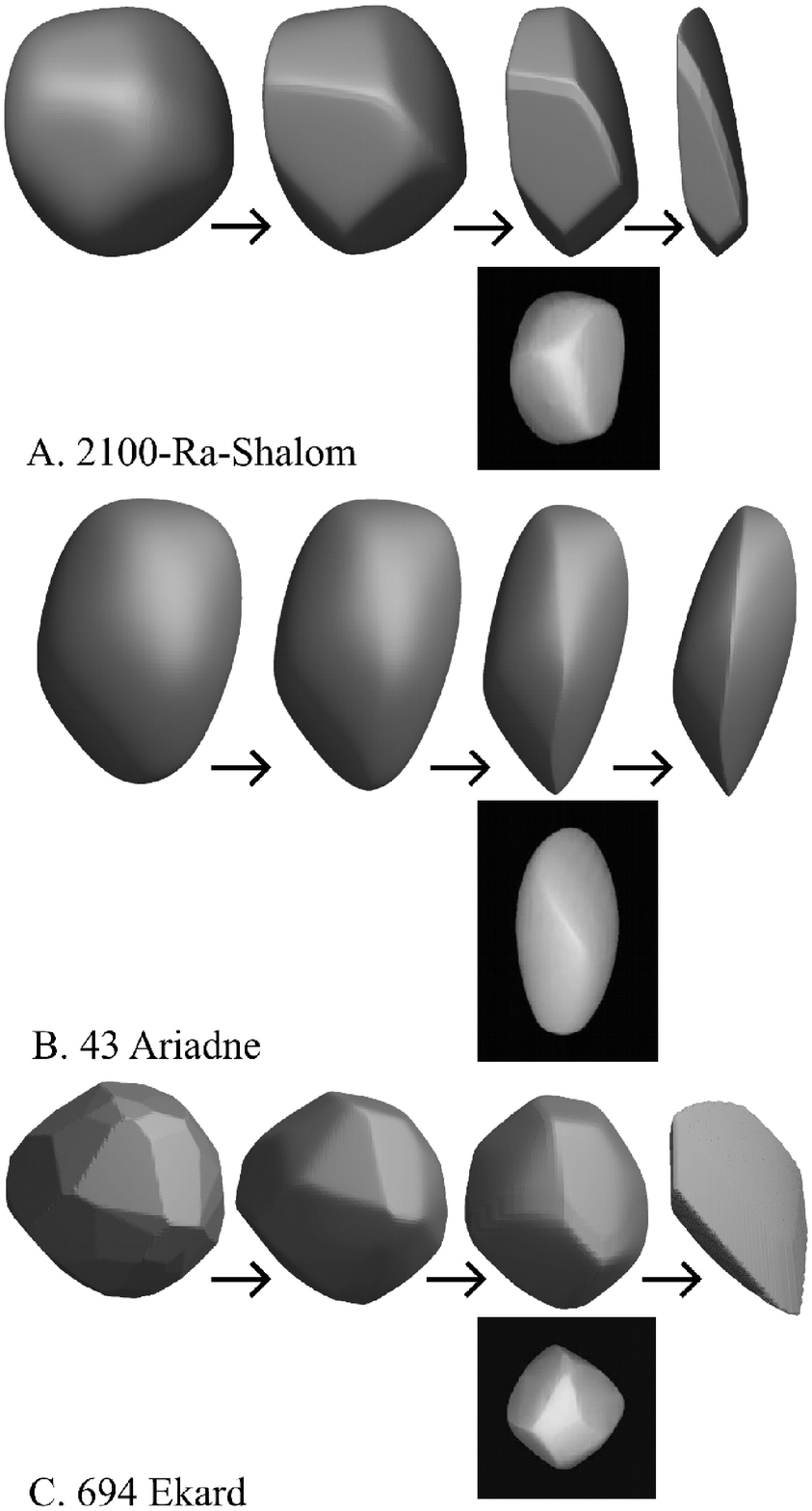}
\caption{Some results of simulated shape evolution by abrasion.
The examples show \
2A: elongated polyhedron, compared to asteroid 2100 Ra-Shalom
(Kaasalainen et al. 2004)
   2B:  flat disk, compared to asteroid 43 Ariadne (KTP2002)
         2C:  polyhedron, compared to asetroid 694 Ekard (Troppa et al. 2004)
        Photometric shape models are shown just to indicate qualitative
agreement.} \end{center} \end{figure}

         \begin{figure}
         \begin{center}
         \includegraphics[width=7.5cm]{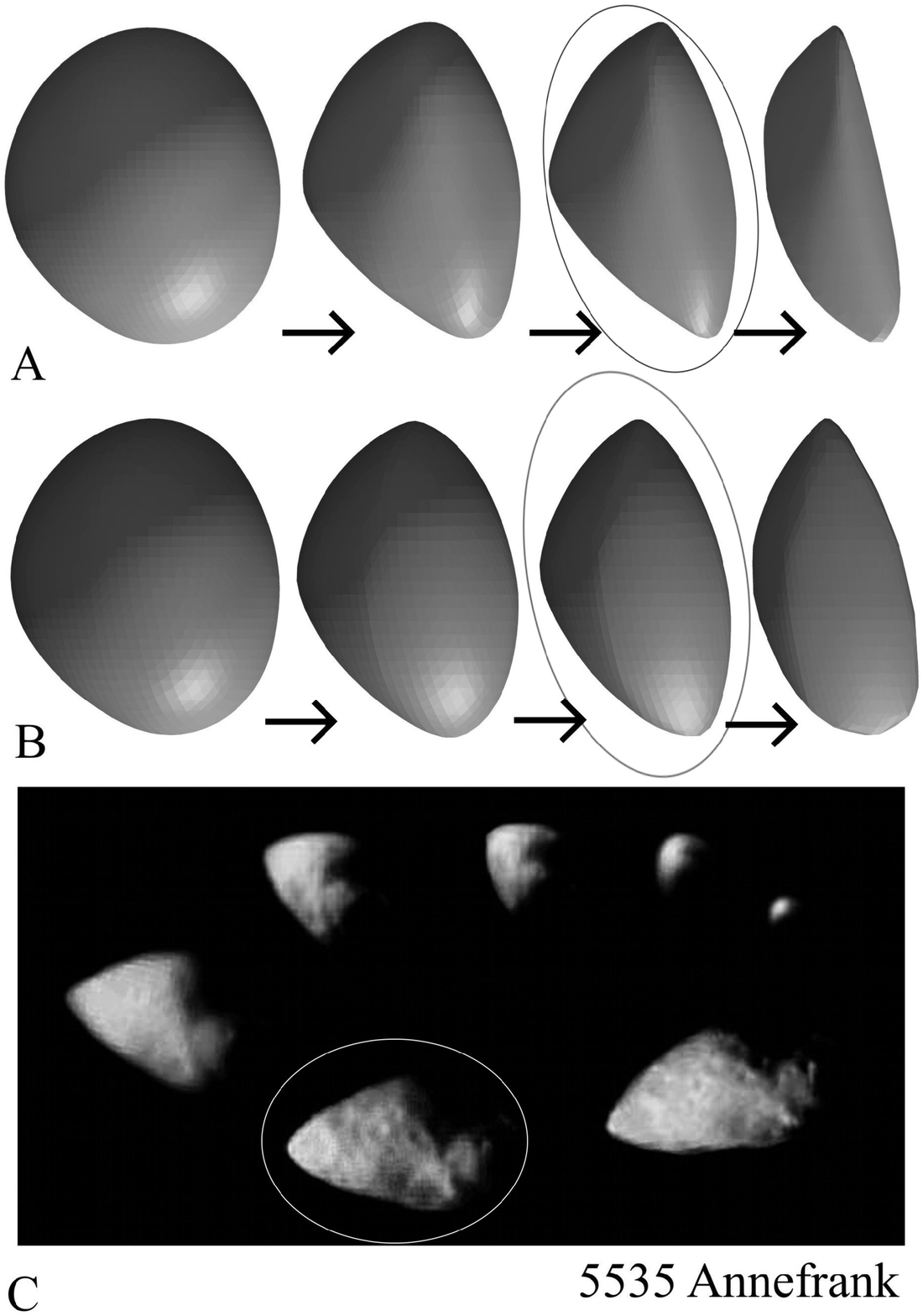}
         \caption{Simulation for Asteroid 5535 Annefrank.
        3A: Hypothetical evolution of the shape under the modeled abrasion
process. In phase 3
        the shape shows remarkable similarity to 5535 Annefrank.
        3B: Hypothetical evolution of the shape under the
            unified model, with 25\% of particles
            captured by gravity.
        3C: Asteroid 5535 Annefrank as seen from the Stardust spacecraft
(small images)}
                 \end{center}
         \end{figure}

So far, all simulations were based on our extreme assumption that all abraded 
material escapes the 
gravitational field. To check the robustness of the described
qualitative features we included in our PDE model the effect
of ejecta falling back to the surface. We integrated the
gravitational potential $U$ for the evolving shape and assumed that 
some fraction $p$ of ejecta resettles to the surface 
with intensity proportional to $e^{U/\overline{U}}$ which is a maximum
entropy approximation of the fallback rate.
Here, $\overline{U}$ denotes the average value of $U$ on the surface. 
All our previous computations correspond to the case $p=0$, whereas
the case $p=1$ represents the scenario
described in more detail by (KA2003). Detailed evaluation of this unified model
is beyond the scope of the current paper, however we
briefly comment on the emerging behavior: for $p=1$ the shapes converge to a sphere
(which would be perturbed if we considered rotational effects);
for smaller $p$, the final shape is a round 
disk with two smooth sides surrounded by a sharp edge;
the flatness of the disk is decreasing function of $p$. 
Figure 3b shows an example with $p=0.25$, illustrating another
 remarkable feature of the process:
 the initial behavior resembles the case $p=0$ (Figure 3a), 
and the emerging polyhedral shape remains visible for a long time
before being transformed to a disk.


                         \section{Conclusions}

We showed in this paper that a simple continuum model, based solely on the
assumptions that asteroids have

\begin{itemize}
         \item{} numerous, high-speed collisions with

         \item{} relatively small impactors
         \item{} arriving uniformly randomly from all directions

         \end{itemize}

accounts for the appearance of flat areas, separated by edges and vertices,
which is observable in the available data.
Photometric imaging certainly does not provide sufficient
information to quantify the uncertainities associated with the
mentioned qualitative geometric features. Based on our present model we can
conclude that asteroids do 
not necessarily evolve preferentially toward regular spheroids, 
but may (depending on the fall-back rate of ejecta)
follow an alternative path, thus confirming 
photometry-derived features. As far as we know, ours
is the only scenario reproducing these features.

Our simple model does not include several effects which
also influence the abrasion.
(For example, collisions with larger objects have totally
different geometric effects: vertices and edges will be worn and appear
less pronounced, craters are created, etc.). The global effects listed
in Sec. 1 also play an important role in forming the final geometry.
Quantitative predictions of shape evolution could only be achieved by more
sophisticated, discrete models. After including the effect of gravity, 
as long as the majority of particles escape, edges remain a visible geometric feature.

The relevance of our model relies on the assumption that high-speed
collisions with small impactors dominate the abrasion process.
This is quite plausible, since the size distribution of asteroids (e.g.
Tedesco et al. 2005, Parker et al. 2008) in the $>500$ m range was found to
be a quickly decreasing function following a power law of exponent
$-2.5\ldots -3$ The size distribution of smaller particles has not been
examined directly, however numerical models for impact evolution predict
even steeper (up to the $-7$ power) size distribution functions especially
for small sizes (Michel et al. 2004). While it is beyond the scope of the
current paper, it is possible to estimate an order of magnitude for the number of 
collisions needed to model the body (KA2003).

Abrasion processes governed by equation (2) occur also on much smaller
scale: pebbles shaped by wind-blown sand , also known as ventifacts, evolve
in a similar manner (V\'arkonyi et al., 2008, Domokos et al., 2009), 
the details of which process we are currently investigating.

\acknowledgments Supports from OTKA grant T72146 (GD, AS, PV),
OTKA K 76816 (GyMSz) and the
``Bolyai J\'anos'' Research Fellowship of the Hungarian Academy of Sciences
(GyMSz) are gratefully acknowledged. The comments by the anonymous Referee 
greatly improved this paper.

\end{document}